\documentclass[11pt,twocolumn]{article}

\usepackage[a4paper,top=3.5cm,width=15.5cm,vscale=.75,head=20.3pt]{geometry}	

\usepackage{graphicx}
\usepackage{dcolumn}
\usepackage{bm}
\usepackage{amsmath,amsfonts}

\usepackage{url}
\usepackage{hyperref} 
\usepackage{multirow}

\usepackage[natbibapa]{apacite}
\let\cite\citep
\usepackage{color}

\begin{document}

\title{Multiphonic modeling using Impulse Pattern Formulation (IPF)}

\author{Simon Linke\textsuperscript{1),2)}, Rolf Bader\textsuperscript{1)}, Robert Mores\textsuperscript{2)}\\
\textsuperscript{1)}Institute of Systematic Musicology,  University of Hamburg, 20354 Hamburg, Germany\\
\textsuperscript{2)}Hamburg University of Applied Sciences, 22081 Hamburg, Germany\\
 linke@mt.haw-hamburg.de}

\twocolumn[
  \begin{@twocolumnfalse}
    \maketitle
    
    \begin{abstract}
      Multiphonics, the presence of multiple pitches within the sound, can be produced in several ways. In wind instruments, they can appear at low blowing pressure when complex fingerings are used. Such multiphonics can be modeled by the Impulse Pattern Formulation (IPF). This top-down method regards musical instruments as systems working with impulses originating from a generating entity, travel through the instrument, are reflected at various positions, and are exponentially damped. Eventually, impulses return to the generating entity and retrigger or interact with subsequent impulses. Due to this straightforward approach, the IPF can explain fundamental principles of complex dynamic systems. While modeling wind instruments played with blowing pressures at the threshold of tone onset, the IPF captures transitions between regular periodicity at nominal pitch, bifurcations, and noise. This corresponds to behavior found in wind instruments where multiphonics appear at the transition between noise and regular musical note regimes. Using the IPF, complex fingerings correspond to multiple reflection points at open finger holes with different reflection strengths. Multiphonics can be modeled if reflection points farther away show higher reflection strength and thus, disrupt periodic motion. The IPF can also synthesize multiphonic sounds by concatenating typical wind instrument waveforms at adjacent impulse time points. \bigskip
    \end{abstract}    
  \end{@twocolumnfalse}
]

\section{\label{sec:1} Introduction}

Single reed instruments (e.g., clarinets or saxophones) can be described as tubes with an attached mouthpiece. An inlet at the mouthpiece is partially covered by a reed and thus forms a valve. Usually, the reed is a thin and elastic piece of cane wood. Several tone holes are drilled into the tube. Opening or closing them changes the effective acoustic length of the tube, and thus the pitch \cite{Fabre.2018}. The flow of air through the tube, which results from blowing into the mouthpiece, usually cannot be investigated analytically \citep[p.401]{Fletcher.2010}. Thus experimental or numerical methods are used.

In the past, many theoretical descriptions of the tone production of clarinets had been given (e.g., \cite{Keefe.1990, Dalmont.1995, Hirschberg.1996}). A first, simple, idealized model system was described by \citet{McIntyre.1983}: Responsible for the movement of the flexible reed is solely the difference between the pressure inside the mouth $p_M$ and the pressure inside the mouthpiece $p_P$. $p_M$ is usually very high, while the airstream entering through the small slit between the mouthpiece and the reed results in low pressure due to Bernoulli's principle. Hence, the reed moves toward the mouthpiece, and the inlet gets closed, resulting in a single pressure impulse entering the tube. In real instruments, the inlet only gets partly closed, as the entering airstream forms a turbulent eddy inside the mouthpiece \cite{Bader.2008}. Thus, the laminar flow inside the instrument is stopped due to turbulent damping. Further, friction may dominate (and Bernoulli's equation should not be used) \cite{Fabre.2018}. 

The pressure impulses entering the mouthpiece can be assumed to be narrow and similar to a Dirac delta function. This impulse travels along the tube and gets partly radiated at the end of the tube or the first open finger hole. However, due to the impedance mismatch between tube and free-field, a considerable amount gets reflected and travels back towards the mouthpiece. Now, the narrow impulse is widened due to damping and inverted due to the reflection at the open end of the cylindrical tube. While returning to the closed reed, the impulse just gets reflected. After a second round trip, the impulse is inverted again and has approximately its initial strength. Thus, the pressure difference between $p_M$ and $p_P$ gets negligible small. The reed returns to its initial position, opening the inlet again. Thus, a new impulse enters the tube.

\textit{"Thus the basic principle of the clarinet functioning can be reduced to a single delayed feedback loop closed on a nonlinear function"} \citep[p. 36]{Rodet.1999b}. Therefore, according to \citet{Fletcher.1979}, highly nonlinear behavior of the generator is crucial. This nonlinearity has been measured several times in the past, first by \citet{Backus.1963} and more recently by \citet{Dalmont.2003}. Furthermore, \citet{Maganza.1986} showed how chaotic behavior can occur in clarinet-like systems when they are excited by nonlinear feedback loops. \citet{McIntyre.1983} introduced a first simple time-domain model of reed instruments only by applying nonlinear excitation to a linear resonator. This approach has been extended by \citet{Taillard.2010}, who describe clarinet-like systems using a nonlinear recursive equation, as well as \citet{Rodet.1999,Rodet.1999b}, who describe several different instruments using feedback loops. 

In reed instruments, there are some technics to produce sounds with more than one harmonic overtone spectrum \citep[p. 226]{Bader.2013}. These so-called multiphonics are played at very high or low blowing pressure \cite{Keefe.1991}. Often combined with uncommon fingerings \cite{Backus.1978}, where the first open finger hole no longer determines the length of the tube and thus the fundamental frequency. When producing multiphonics, complex patterns of open and closed holes are used. As a result, not a single impulse returns to the reed. Instead, a complex time-dependent pressure disturbs the reed.

Under these conditions, the embouchure and the blowing pressure must be controlled very carefully \cite{Bader.2013}. Further, the vocal tract can have a crucial impact on the perceived frequencies and their amplitudes \cite{Chen.2011}. Due to the many degrees of freedom when producing multiphonics, modeling is not straightforward, and model parameters must be fine-tuned carefully. In this study, it is focused on complex fingerings and low blowing pressure. Here multiphonics can be reliably reproduced as mode-locking happens more likely at high amplitudes \cite{Fletcher.1978,Chen.2011}.


The physics of tone production of single-reed instruments can be described by solving the Navier-Stokes equation. However, this can only be done numerically and requires vast computing power and time. Further, the solution \textit{"would not reveal a great deal about the general principles underlying the mechanism of sound generation and control. It is therefore appropriate to use much simpler models [...]"} \citep[p.401]{Fletcher.2010}. Numerical, physical models (e.g., FEM) are typically suitable for a general, qualitative description of the phenomenon or specialized investigations of a given geometry. While targeting models with a musical focus and optional real-time synthesizing, abstractions and comprehensions are helpful, according to \citet{Rodet.1999}. Musicians must be able to understand and use those models. Thus, one should develop \textit{"models that retain the essence of the behavior of a class of instruments while disregarding all details that are not useful for understanding what is typical of that class"} \citep[p. 18]{Rodet.1999}.

\citet{Taillard.2010} investigate stable tone production of clarinet-like systems using nonlinear recursive equations. They further investigate bifurcating regimes which \textit{"might be related to some kinds of multiphonic sounds produced by the instrument"} \citep[p.268]{Taillard.2010}. This investigation systematically describes multiphonics using a more applicatory recursive equation. The production of multiphonics with a clarinet is modeled using the Impulse Pattern Formulation (IPF), an abstract recursive equation that can model musical instruments in general. The IPF is capable of dealing with the complex geometry of an instrument. It is even possible to model the advanced fingerings essential for multiphonics.

To prove the results, the IPF-model is used to synthesize different multiphonics. The aim is not to achieve a detailed reproduction of a given multiphonic but rather to find the general requirements of the IPF and its system parameters to reproduce a certain class of multiphonics. All numerical calculations are done using the library \textit{``DynamicalSystems"} by \citet{Datseris.2018} for the programming language \textit{``Julia"} (see \cite{Bezanson.2017}).

\section{Impulse Pattern Formulation \label{sec:IPF}}

As mentioned in Section \ref{sec:1} most numerical models of musical instruments are specialized in answering very detailed questions, but they are usually not suitable to derive general physical or musical principles. For this purpose \citet{Rodet.1999,Rodet.1999b} developed simplified models using delayed-feedback loops. Thus they detect typical elements for several classes of sustained instruments. The Impulse Pattern Formulation (IPF) takes this idea one step further: It is a top-down method that describes the transient behavior of arbitrarily coupled systems, and thus, all kinds of musical instruments. As the IPF uses only a limited number of system parameters, it allows, for instance, a straightforward comparison of different instruments. Thus, it can also be used to identify the general structure of a system capable of producing multiphonics, as done in this research.

\subsection{Deriving the fundamental equations \label{sec:derivIPF}}

The IPF was introduced by \citet{Bader.2013} with a focus on tone production of musical instruments. Musical instruments are assumed to consist of nonlinearly coupled subsystems that are excited by impulses. This might be obvious when talking about plucked string instruments or percussion instruments. However, it is also true for any other instrument, e.g., bowed string instrument, where the bow displaces a string until the string suddenly slips back, which results in an impulse traveling along the string and returning to the bow contact point to retrigger the next period (see, e.g., \citet{Cremer.1984} or \citet{Giordano.2018}). As already described in Section \ref{sec:1}, the tone production in single-reed instruments is likewise based on distinct impulses that enter the tube when the reed is open.

\begin{figure*}[htb]
  \begin{center}
    \includegraphics[width=1 \linewidth]{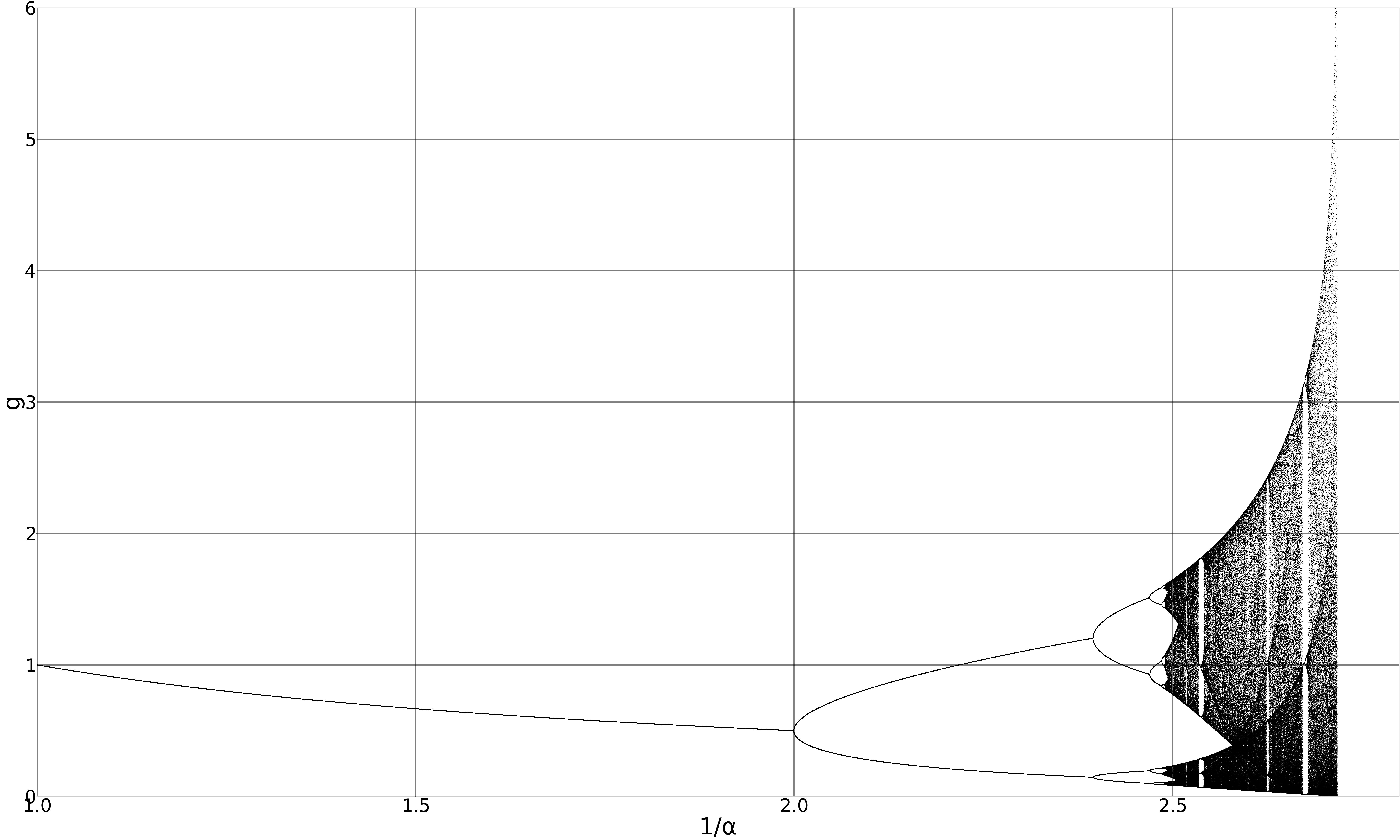}
    \caption{\label{fig:BifSimple}{Bifurcation scenario of the IPF in its most simple form}}
  \end{center}
\end{figure*}

In contrast to classical generator-resonator models (e.g., \cite{Fletcher.1978}), the IPF describes musical instruments as self-organizing systems and takes a more general approach: A musical instrument is a system acting upon itself, consisting of mutually coupled subsystems, possibly even interacting backward. The system can be analyzed from the perspective of any subsystem, as it sends out impulses while responding to other subsystems. While describing single-reed instruments, the most simple modeling uses two subsystems: a reed and a tube. Taking the point of view on the reed, it sends out impulses into the tube. These impulses get reflected at the end of the tube and return to the reed, where they displace the reed, thereby triggering the next impulse. Thus, according to \citet[p. 286]{Bader.2013}, the change of the system is caused by the system itself,

\begin{equation}
  \frac{\partial \bar g}{\partial t}=\frac{1}{\alpha} \bar g ,
\end{equation}

where $\bar g$ is the system state. It reflects both the amplitude and the periodicity. However, the precise meaning depends on the observed system. $\alpha$ represents the strength of the back traveling impulse, which is determined by the reflection strength of the second subsystem. The strength of the back traveling impulse also highly depends on the input strength of the system. Thus, $\alpha$ also refers to the blowing pressure when describing single-reed instruments. Explanations for single-reed instruments are given in Sections \ref{sec:MultiModel} and \ref{sec:Synthese}. \citet{Linke.2019c} discuss applications for different musical instruments.

The Impulse needs a certain amount of time (one period $T_0$ of the fundamental frequency $f_0$) to travel to the end of the tube and back again during which the impulse gets exponentially damped.  Talking this into account, \citet[pp. 286-288]{Bader.2013} deduces the IPF in its most simple Form,
\begin{equation}
  g_+=g-ln \left( \frac g \alpha \right) \label{eq:IPFsimple} ,
\end{equation}

where $g$ is the system state at a certain time step and $g_+$ is the succeeding system state. There is no precise time interval between $g$ and $g_+$. It is the time until a new event occurs. When modeling musical instruments, this is $T_0$. Choosing an initial value for $g_0$, Eq. \eqref{eq:IPFsimple} can be iteratively calculated. After several iterative steps, the IPF can diverge or converge to a limit, depending on $\alpha$. Further, the IPF can show chaotic behavior like bifurcations. If $\alpha$ is constant the IPF usually converges after $n < 300$ iteration steps. To investigate the bifurcation scenario of the IPF, 2500 iteration steps are performed, and the last 250 values $g_n$ are plotted in Figure \ref{fig:BifSimple} to illustrate dependency on alpha. This is done using the function "orbitdiagram" from the library \textit{"DynamicalSystems"} \cite{Datseris.2018}. 

Low values $1/\alpha$, which refer to high blowing pressures, result in stable behavior of the IPF. As only one $g$ is plotted for every alpha, the IPF converges to a stable limit. Above $1/\alpha = 2$, the limits suddenly split into two branches. Here, two alternating limits occur, and two system states can coexist. Higher values $1/\alpha$ lead to higher-order bifurcations and finally to a chaotic domain, where an infinite number of (unstable) system states exist. This chaotic region is irregularly interrupted by small regions of countable system states.

If the instrument geometry is more elaborated or increasingly complex fingerings are used, more subsystems and thus more reflection points must be considered. The sent-out impulses will return at subsequent time steps and related specific strength $\beta_k$. Thus, the following system state $g_+$ is influenced by the system states of earlier time points $g_{k-}$. Based on these assumptions, the IPF is derived in its general form \citep[pp. 290-291]{Bader.2013} :

\begin{equation}
    g_+=g-ln \left( \frac 1 \alpha \left( g-\sum^{n}_{k=1} \beta_k e^{g-g_{k-}}\right)\right) 
    \label{eq:IPF}
\end{equation}

\begin{figure}[ht]
  \begin{center}
    \includegraphics[width=1 \linewidth]{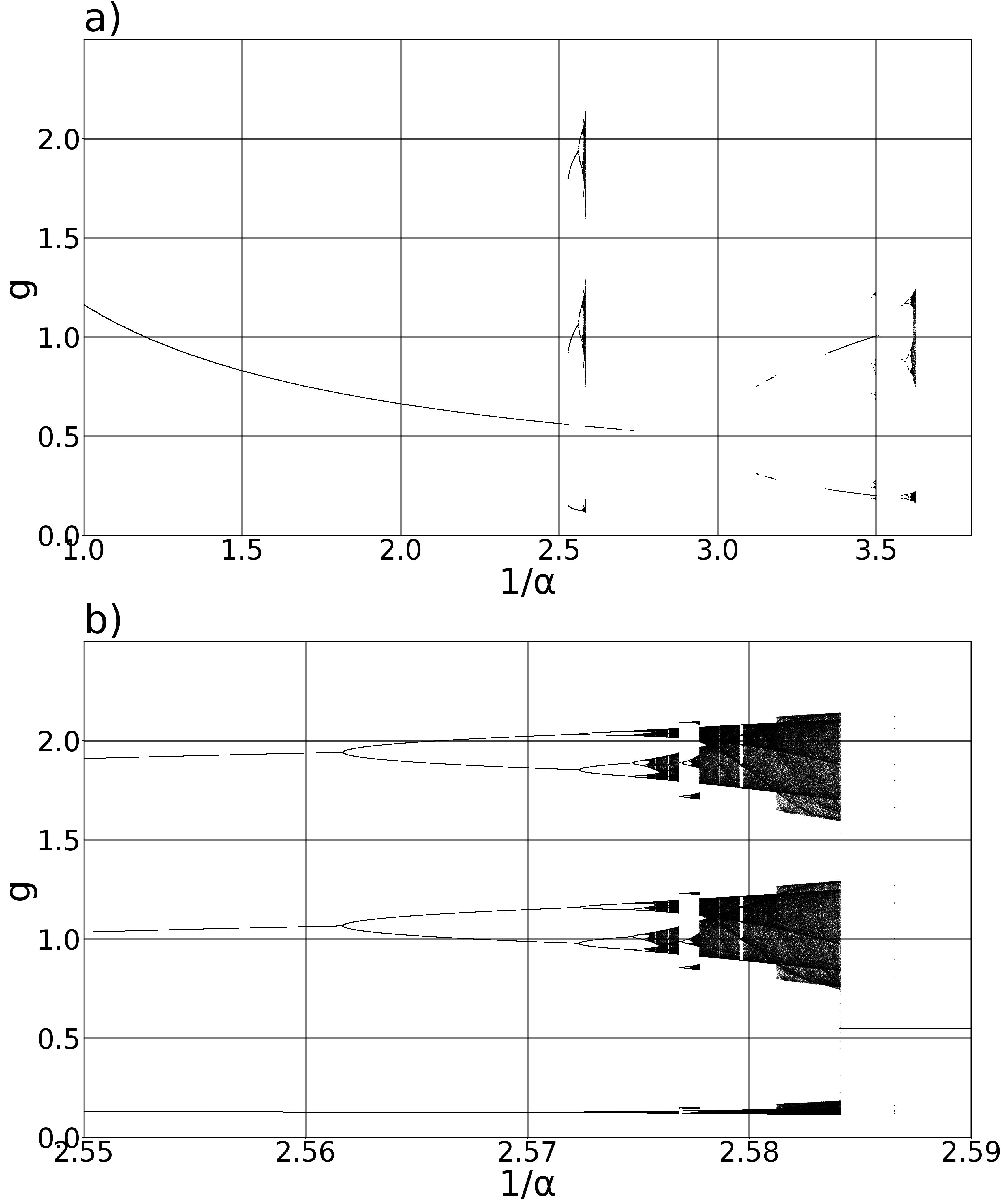}
    \caption{\label{fig:BifBeta} {Bifurcation scenario of the IPF with two reflection points in dependence of $1 / \alpha$ with $\beta=0.164$. The lower chart \textbf{b)} is a zoom of the chaotic region.}}
  \end{center}
\end{figure}

Depending on $\beta_k$, the IPF can become quite complex. The bifurcation scenario can be similar to the one described in Figure \ref{fig:BifSimple}. However, adding just one single reflection point can change this significantly. Figure \ref{fig:BifBeta} shows the bifurcation scenario for a single reflection point $\beta=0.164$. Due to the exponential function in Eq. \eqref{eq:IPF} two initial values $g_{01}$ and $g_{02}$ must be assumed. When choosing $g_{01}=0.3$ and $g_{02}=0$, there is no straight transition from stable, via chaotic to diverging states, directly. It is possible to return from a chaotic stage to a stable state. Also, stable and chaotic domains can be interrupted by small diverging regions. 

\subsection{Boundaries and stability of the IPF}

In Section \ref{sec:MultiModel} the IPF is used to model multiphonics. Even though the underlying equation of the IPF is quite simple, it is not trivial to find an appropriate set of system parameters for modeling multiphonics. It is essential to reduce the set of possible parameters for this multi-dimensional system. Therefore, some observations of the mathematical limits of the IPF are helpful to determine the different stable or chaotic domains and identify the sparse parametric range capable of producing multiphonics reliably. Since a suitable combination is found and is used for sound synthesis like done in Section \ref{sec:Synthese}, it is also crucial to know the boundaries of the suitable parameter range. A comprehensive investigation on the parametric space, the boundary conditions, as well as convergence and stability, has been done by \citet{Linke.2019b} and its main findings are summed up in the following section.

The parameter $\alpha$ is related to the input strength of the system, the blowing pressure. Thus, only positive values $\alpha$ are physically reasonable. As the reflection strengths, $\beta_k$ have to be positive, too. \citet{Linke.2019b} derive a first restriction referring to the law of conservation of energy:

\begin{equation}
  \alpha \geq \sum^n_{k=1} \beta_k \label{eq:Energy}
\end{equation}

According to \citet[p. 291]{Bader.2013}, higher orders $k$ will result in smaller $\beta_k$, as they represent reflection points farther away from the excitation point. Therefore, reflected impulses return later and weaker. Hence, there is a second condition for the relationship of $\alpha$ and $\beta_k$:   
\begin{equation}
\alpha > \beta_1 >\beta_2 >\beta_3 > ... >\beta_n\label{eq:Kaskade}  ,
\end{equation}
which is at least valid for similar transitions from one subsystem to another (e.g. similar impedance differences between the subsystems). In Section \ref{sec:multiCla} it is shown, that this rule may not be relevant for all musical instruments in general, as small perturbations are possible even close to the excitation point.


Referring to Figure \ref{fig:BifSimple}, parameter $g$ reaches no values above $1/ \alpha \approx 2.7$. In this region, $g$ becomes complex and starts to diverge. Talking about free reed instruments, it is evident that no tone production is possible if no one blows into the mouthpiece. Consequently, there seems to be a minimum $\alpha_{min}$, which is the lower boundary for physically reasonable behavior. One only obtains complex values $g_+$ from Eq. \eqref{eq:IPF} if the argument of the logarithm is negative. It has already been shown that it is not possible to derive a general limit by analytical means, as the solution depends on the initial values \cite{Linke.2019b}. This is only possible if just one reflection point exists ($\beta_k=0 \; \forall k$):

\begin{equation}
  \frac 1 \alpha_{min} =e \approx 2.7 \label{eq:mina}
\end{equation}
As soon as one or more $\beta_k$ occur, $\alpha_{min}$ has to be calculated numerically. 

There is a fixed point $f(g_s)=g_s$ of the IPF, where the system state $g$ is constant for all iteration steps:
\begin{equation}
g_-=g=g_+=g_{2+}=\ldots=g_{n+}=g_s
\end{equation}
Thus, simplifying Eq. \eqref{eq:IPF} leads to a fixed point:
\begin{align}
    \nonumber
    g_s&=g_s-ln\left(\frac 1 \alpha \left(g_s-\sum^{n}_{k=1} \beta_k e^{g_s-g_s}\right)\right)\\ 
    g_s&=\alpha+\sum^{n}_{k=1} \beta_k \label{Eq:Konvergenzpunkt}
\end{align}

A fixed point $g_s$ exists for any combination of $\alpha$ and $\beta_n$. But, as \citet[pp.65-66]{Argyris.2015} stated for recursive differential equations in general, it is only stable if the absolute value of the derivative $\partial g / \partial \alpha <1$. Again, \citet{Linke.2019b} show that no general analytical solution exists, due to the dependence on initial values. The equation can be solved analytically if one reflection point exists:
\begin{equation}
  \alpha>0.5
\end{equation}

The critical point $\alpha_c=0.5$ is called the first bifurcation point. It marks the transition from stable states to bifurcation as already observed in Figure \ref{fig:BifSimple}.

Finally, increasing the number of reflection points usually increases the system's stability. Due to the term $e^{g-g_{k-}}$ in Eq. \eqref{eq:IPF} not every $\beta_k$ has an inevitable impact at every instant. Thus, Eq. \eqref{eq:Energy} can be slightly violated. Then, bifurcation hardly occurs. Thus, this domain is still negligible when modeling multiphonics. \cite{Linke.2019b}

\section{Modelling multiphonics}\label{sec:MultiModel}

In this section, the IPF is used to model multiphonics of a clarinet. It is focused on B$\flat$ clarinets, with keys and holes arranged according to the Boehm system. The set of possible multiphonics is to reduce to those with just two perceived pitches. Further, the multiphonics are playable without additional technics like, e.g., flutter tongue, thrills, or tremolos.

\subsection{IPF model of a clarinet}\label{sec:ClariModel}

Stable tone production of single-reed instruments can be described by the IPF in its most simple form as given by Eq. \eqref{eq:IPFsimple}. There is just one reflection point at the end of the tube (the first open finger hole), which is represented by $\alpha$. As mentioned above, it is possible to add reflection points by opening additional finger holes. This adds further $\beta_k$ to the IPF. 

\begin{figure*}[ht]
  \begin{center}
    \includegraphics[width=.66\linewidth]{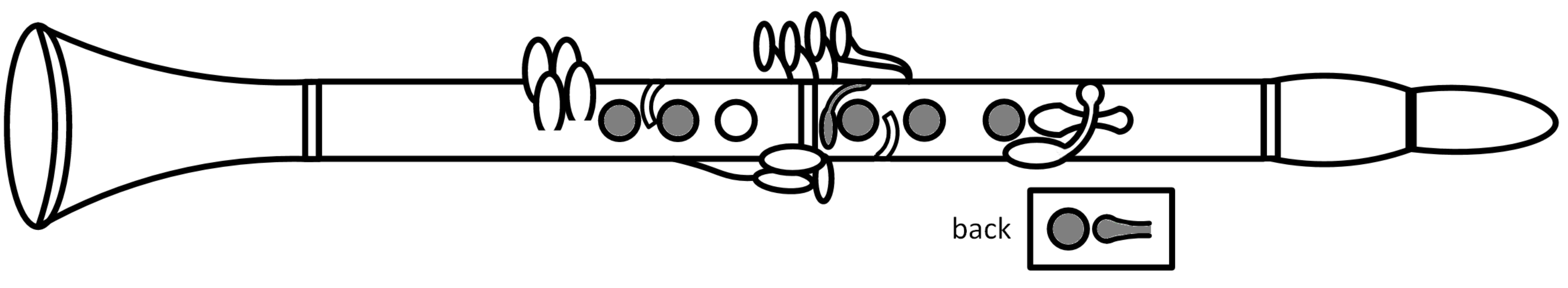}
    \caption{\label{fig:MultiphonicFingering}{Sketch of a B$\flat$ clarinet (Boehm system): gray circles correspond to closed holes when producing a multiphonic containing the notes $E_4$ and $G_5$}}
  \end{center}
\end{figure*}

Figure \ref{fig:MultiphonicFingering} shows a fingering for a multiphonic played on a clarinet, which can be found in different collections of multiphonic fingerings \cite{Roche.2018,Farmer.1977}. The resulting sound is described to consist of two dominating frequencies, which approximately could be notated as the musical interval $E_4$-$G_5$ (15 semitones). There are three regions of open holes which are likely to reflect the sound wave propagating through the tube. They can be transferred to the IPF. The first reflection point occurs due to the open register key on the back. It lies close under the barrel-joint and is represented by $\alpha$. The pressed C$\sharp$-key results in another open hole on the instrument's back. It lies close to the open hole in the middle of the instrument. So, both could be condensed to one reflection point $\beta_1$. The effective open end of the tube is given by two open holes between the undermost key and the bell and is represented by $\beta_2$. So, the resulting equation for this fingering is
\begin{equation}
  g_+=g-ln\left(\frac{g-\beta_1 e^{g-g_{-}}-\beta_2 e^{g-g_{2-}} }{\alpha}  \right) .
\end{equation}

$\beta_1$ and $\beta_2$ are introduced as reflection strengths. They not only represent impedance differences inside the tube. They are meant to represent the impact that returning impulses have on the reed. Thus, they also take care of constructive or destructive interferences due to path differences.
Furthermore, the IPF with three reflection points needs three initial values, which can be reduced to one initial value $g_0$, when calculating the other necessary values with a simplified version of the IPF as described by \citet[p. 6]{Linke.2019b}.

The approach described above is chosen as simple as possible. As stated by \citet{Bader.2013}, it is an advantage of the IPF that the number of observed subsystems can be reduced if a less detailed observation of the overall system is sufficient to answer a specific research question (for a detailed description, see \cite{Linke.2021c}). In real instruments, all open finger holes and also closed finger holes can cause weak reflections (see, e.g., \cite{Fabre.2018,Nederveen.1969}). They can be considered at additional reflection points when, e.g., an in-depth comparison of certain multiphonics is performed.

While modeling multiphonics with the IPF, it is plausible that these relate to bifurcating regions, as $g$ was supposed to be related to the periodicity.
Multiphonics with two dominating frequencies correspond to first order bifurcations in the orbit diagramm (e.g. in Figure \ref{fig:BifSimple} $1/\alpha \in [2, 2.4]$). Hence, the audible interval $I$ is represented by the ratio of the possible system states 
\begin{equation}
  I=\frac{g_k}{g_{k-}} . \label{eq:modF}
\end{equation}

Thus, the multiphonic shown in Figure \ref{fig:MultiphonicFingering}, which represents a musical interval of 15 semitones, is related to the ratio $g_k / g_{k-}\approx2.38$, respectively the reciprocal $g_{k-} / g_{k}\approx 0.42$.  In the following sections, all intervals are calculated to be contained in the Intervall $]0,1]$ to allow a more intuitive comparison of different multiphonics.

Higher-order bifurcations represent multiphonics with more than two dominating frequencies. Thus, more consecutive system states $g_k-$ have to be considered for the audible interval. Stable regions of the IPF are respectively related to stable tone production and chaotic regions to noisy sounds.

\subsection{Determining modelling parameters}\label{sec:multiCla}

\begin{figure}[ht]
  \begin{center}
    \includegraphics[width=.98 \linewidth]{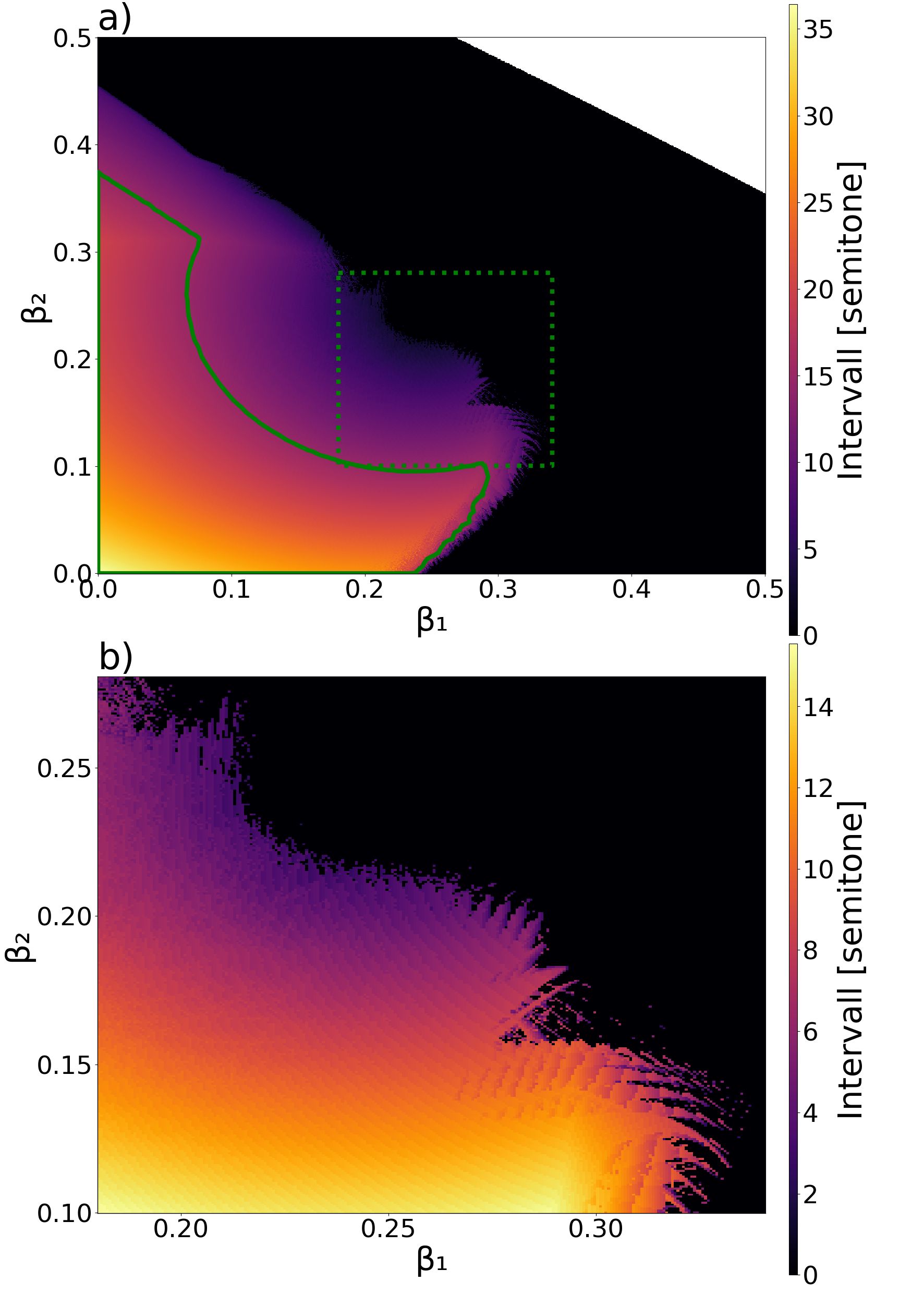}
    \caption{\label{fig:IntervallsPossible} {Maximum audible interval when modelling a multiphonic with a given combination of $\beta_1$ and $\beta_2$. In black regions only stable tone production is possible and in white regions no tone production is possible at all. Inside the solid green outlined area, the muliphonic shown in Figure \ref{fig:MultiphonicFingering} can be modeled. b) shows a zoom of the dotted green rectangle shown in a) }}
  \end{center} 
\end{figure}

The IPF is used to model a variety of multiphonics playable on the B$\flat$ clarinet (Boehm system) and similar to the one introduced in Figure \ref{fig:MultiphonicFingering}: Two perceived pitches and not more than three possible reflection points $\alpha$, $\beta_1$, and $\beta_2$. Therefore, collections of possible multiphonics of \cite{Farmer.1977}, \cite{Roche.2014} and \cite{Roche.2018} are considered to find 236 suitable multiphonics. The audible intervals range from 0.5 to 26.5 semitones. Different multiphonics produce the same but transposed interval. Thus, the set can be reduced to 53 different intervals. The IPF is calculated for a range of $\beta_1$ and $\beta_2$. Thus, it can be explored how the IPF's dimensional space relates to the population of multiphonics. $\alpha$ is varied between 0 and 1 and the maximum maximum interval $g_k / g_{k-}$ is shown in Figure \ref{fig:IntervallsPossible}. It is always possible to produce smaller intervals, at least for several optimized initial values $g_0$, which is plausible when looking at Figures \ref{fig:BifSimple} and \ref{fig:BifBeta}.

All playable intervals can be modeled using the IPF. Considering that only the maximum intervals are plotted, there is a broad region of combinations of $\beta_1$ and $\beta_2$ for which it is possible to produce a desired multiphonic, especially the smaller intervals. For instance, the multiphonic shown in Figure \ref{fig:MultiphonicFingering} can be modeled with any combination of $\beta_1$ and $\beta_2$ inside the solid green outlined area. Even though the overall shape seems to be quite regular, it can be observed that the structure is chaotic and shows self-similarities, Figure \ref{fig:IntervallsPossible} b).

The indicated regions to produce multiphonics are broad, and further assumptions can be made to identify more specific parameter combinations which are more likely to produce multiphonics. As there are no further mathematical restrictions, these assumptions can only be made from educated musical praxis: Multiphonics can reliably be produced by skilled players and for an extended period. During tone production, a musician constantly listens to the produced sound and slightly adapts the blowing pressure \cite{Roche.2014}.

Therefore small changes in blowing pressure should not change the audible frequencies significantly. Thus, the derivative of the interval $I$ with respect to the blowing pressure
\begin{equation}
  \frac{\partial I }{\partial \alpha} =\frac{\partial\left( \frac{g}{g_-}\right) }{\partial \alpha} 
\end{equation}
should be as small as possible. Furthermore, for reliable tone production, it should be possible to model a particular multiphonic independent of the initial value $g_0$. Thus, the IPF is calculated for 150 different initial values between $0$ and $5$ to achieve a probability for a reliable reproduction. Dividing this probability by the derivative mentioned above leads to a likelihood for producing a certain multiphonic at a given parameter combination.

\begin{figure}[ht]
  \begin{center}
    \includegraphics[width=.99 \linewidth]{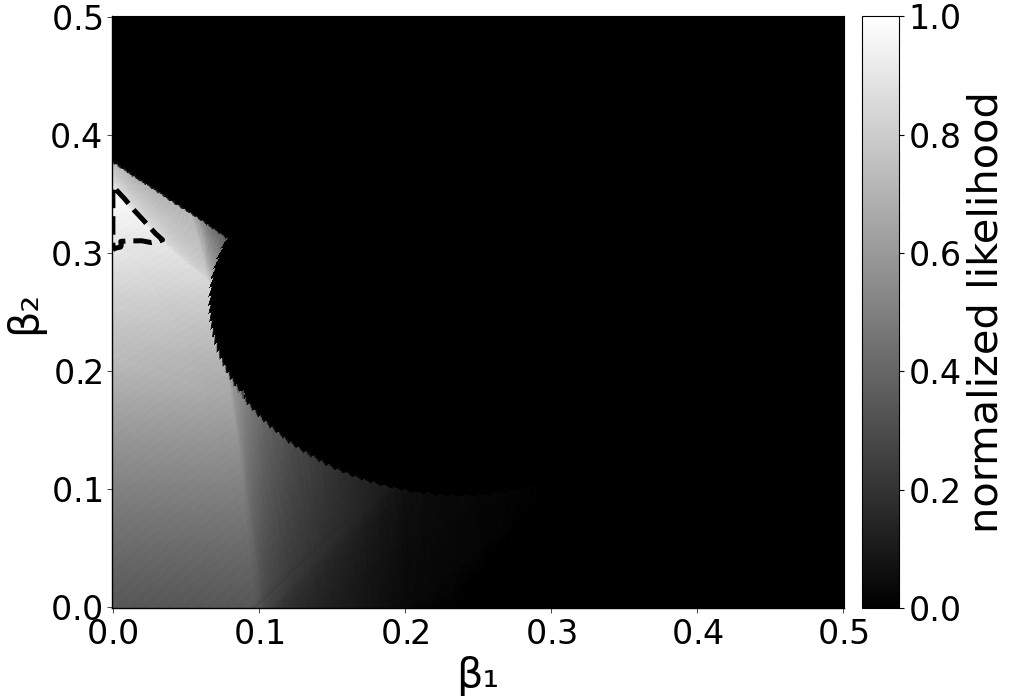}
    \caption{\label{fig:Likelihood} {Normalized likelihood for producing the multiphonic shown in Figure\ref{fig:MultiphonicFingering}. The dashed outlined area shows a normalized likelihood of 0.9 or above}}
  \end{center}
\end{figure}

\begin{figure*}[ht]
  \begin{center}
    \includegraphics[width=.66 \linewidth]{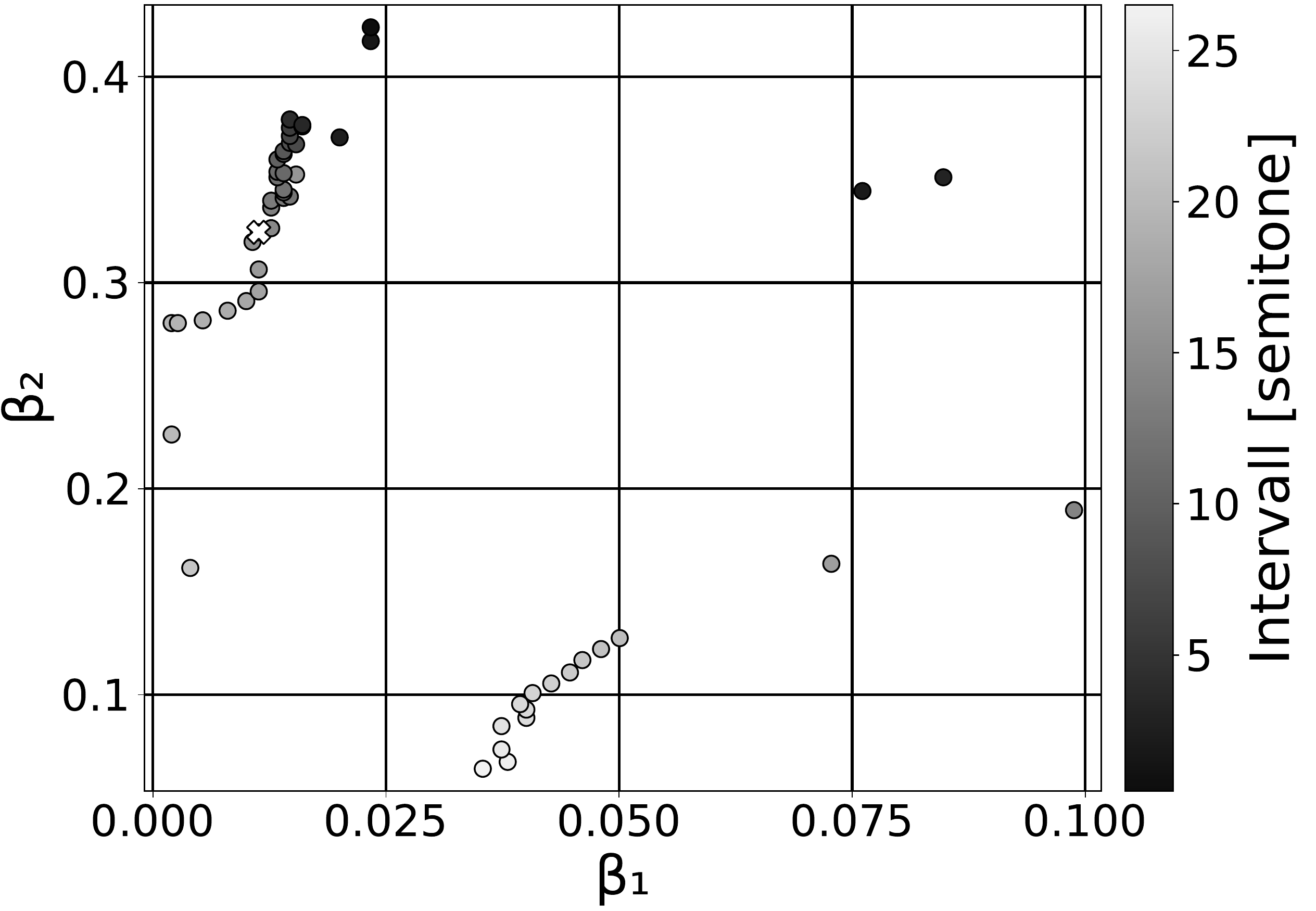}
    \caption{\label{fig:Schwerpunkt} {Most likely combinations of $\beta_1$  und $\beta_2$ for producing the observed multiphonics. The multiphonic shown in Figure \ref{fig:MultiphonicFingering} and~\ref{fig:Likelihood} is marked as a white cross.}}
  \end{center}
\end{figure*}

Figure \ref{fig:Likelihood} shows the normalized likelihood for producing the multiphonic shown in Figure \ref{fig:MultiphonicFingering}. A maximum likelihood can be observed for $\beta_1 \in [0, 0.05]$ and $\beta_1 \in [0.3, 0.36]$. This likelihood is calculated for all 53 observed intervals. Every combination which is at least 90~\% of the maximum likelihood is taken into account. The resulting area, which is highlighted in Figure \ref{fig:Likelihood} is integrated to one single point to allow a deep comparison of the modeling parameters of different multiphonics. Therefore, the weighted mean of the area is calculated. 
Here, every combination of $\beta_1$ and $\beta_2$ is weighted by the associated likelihood. The results are shown in Figure \ref{fig:Schwerpunkt}.

In general, the reflection strength at the end of the tube, $\beta_2$, decreases with increasing intervals. Furthermore, $\beta_1$ is generally smaller than $\beta_2$. This seems to be a crucial requirement for producing multiphonics. For stable tone production, the first reflection point dominates (and determines the fundamental frequency). However, when producing multiphonics, the reflection point most far away from the excitation point must be strong enough to disrupt this periodic motion.
The large value of $\beta_2$ is also reasonable when looking at the instrument geometry. The impedance mismatch between tube and free-field at the end of the tube is much stronger than the impedance changes introduced by opening relatively small holes in the middle of the tube.

This ratio of $\beta_1$ and $\beta_2$ violates the condition given by Eq. \eqref{eq:Kaskade}, introduced by \citet{Bader.2013}. But \citet{Bader.2013} motivates the relation focussing on stable tone production. Figure \ref{fig:Schwerpunkt} shows, that it is not valid when focussing on bifurcations.

\section{Sound synthesis} \label{sec:Synthese}

Having investigated the parameters $\beta_1$ and $\beta_2$ for a given multiphonic, the IPF can be used to synthesize the resulting sound. As stated in Section \ref{sec:derivIPF}, the time between two iteration steps is one period $T_0$, the inverse of the fundamental frequency $f_0$. So, the resulting waveform is calculated for every single period concerning the related system state $g$.

First of all, the IPF not only models the final systemic state, be it stable or rather unstable, in the case of multiphonics. It also models the dynamics from the onset and initial transients to a finally developed sound. The dynamics include articulation and dynamic expression of a musician, which can also be modeled by varying $\alpha$ in time through the initial transient phase. However, as the IPF only describes the development of an impulse pattern, it does not provide any spectral information. This is no problem in practice, as we usually know an instrument's spectrum during stable tone production. Further, this allows combining the transient behavior of one instrument with the spectral content of another instrument to achieve new, hybrid instruments like, e.g., a bowed tuba or a plucked flute. In this work, three different spectra are chosen: 
\begin{enumerate}
  \item A Gaussian function as a very simple approximation
  \item One period of a recorded clarinet playing a stable sound as an example of a rather realistic sound 
  \item One period of a bowed string as an example of a mismatching sound
\end{enumerate} 

\begin{figure}[ht]
  \begin{center}
    \includegraphics[width=1 \linewidth]{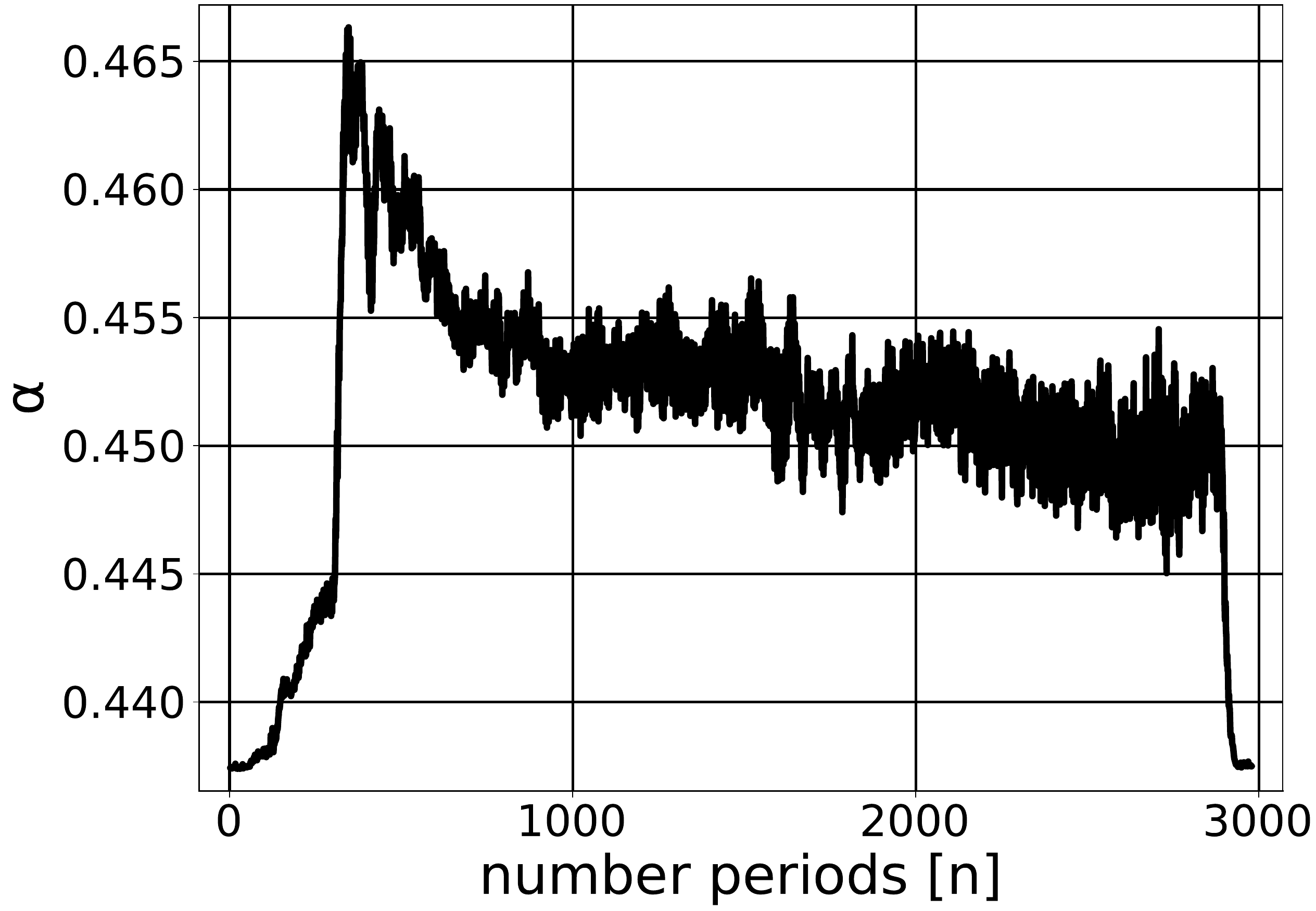}
    \caption{\label{fig:tima} {Scaled enevelope of the recording of a multiphonic to deduce a time series of the control parameter $\alpha$}}
  \end{center}
\end{figure}

Further, an appropriated time series of the blowing pressure $\alpha$ has to be chosen. To achieve a natural, realistic transition, the envelope of a recorded multiphonic (audio example 1 from \cite{Linke.2022} ) is chosen. It is scaled to match the $\alpha$ of the produced multiphonic. The lowest value of the envelope must equal $\alpha_{min}$ like described by Eq. \eqref{eq:mina}. As $\beta_k \neq 0$, it must be calculated numerically. If the envelope reaches a constant level, it should equal the $\alpha$ which is necessary to produce the audible interval $g_k/g_{k-}$ of the investigated multiphonic. A resulting series is shown in Figure \ref{fig:tima}. Using this series of $\alpha$, a time series for the system state $g$ can be calculated using Eq. \eqref{eq:IPF} and inserting $\beta_1$ and $\beta_2$ investigated in Section \ref{sec:multiCla}. 



As stated in section \ref{sec:derivIPF}, one iteration step of the IPF equals one periode of the produced sound. So, for each iteration step one period of the chosen waveform is is concatenated by consecutively copying it into a buffer. According to Eq. \eqref{eq:modF} the length of every period $T$ is changed according to the ratio of the system states:
\begin{equation}
    T=\frac 1 {f_0} \frac {g} {\tilde{g}}
\end{equation}
Choosing $\tilde{g}=g$ results in a signal which oscillates with the fundamental frequency $f_0$. In Section \ref{sec:derivIPF} it was stated that $g$ also refers to the amplitude. So, the amplitude of every period should be modulated by multiplying the related system state $\tilde{g}$.

Now a second signal can be added, where $\tilde{g}$ equals the previous system state $g_-$. Thus, both signals oscillate with the same frequency $f_0$ if $g=g_-$. As soon as bifurcation occurs, two pitches are perceived. Adding additional signals where $\tilde{g}$ equals earlier system states ($g_{2-}$, $g_{3-}$, $g_{4-}$, ...) allows to synthesize higher-order bifurcations or a more precise recreation of noisy states. The overall volume should be further scaled according to the blowing pressure $\alpha$ for a more realistic impression.

\begin{figure}[ht]
  \begin{center}
    \includegraphics[width=1 \linewidth]{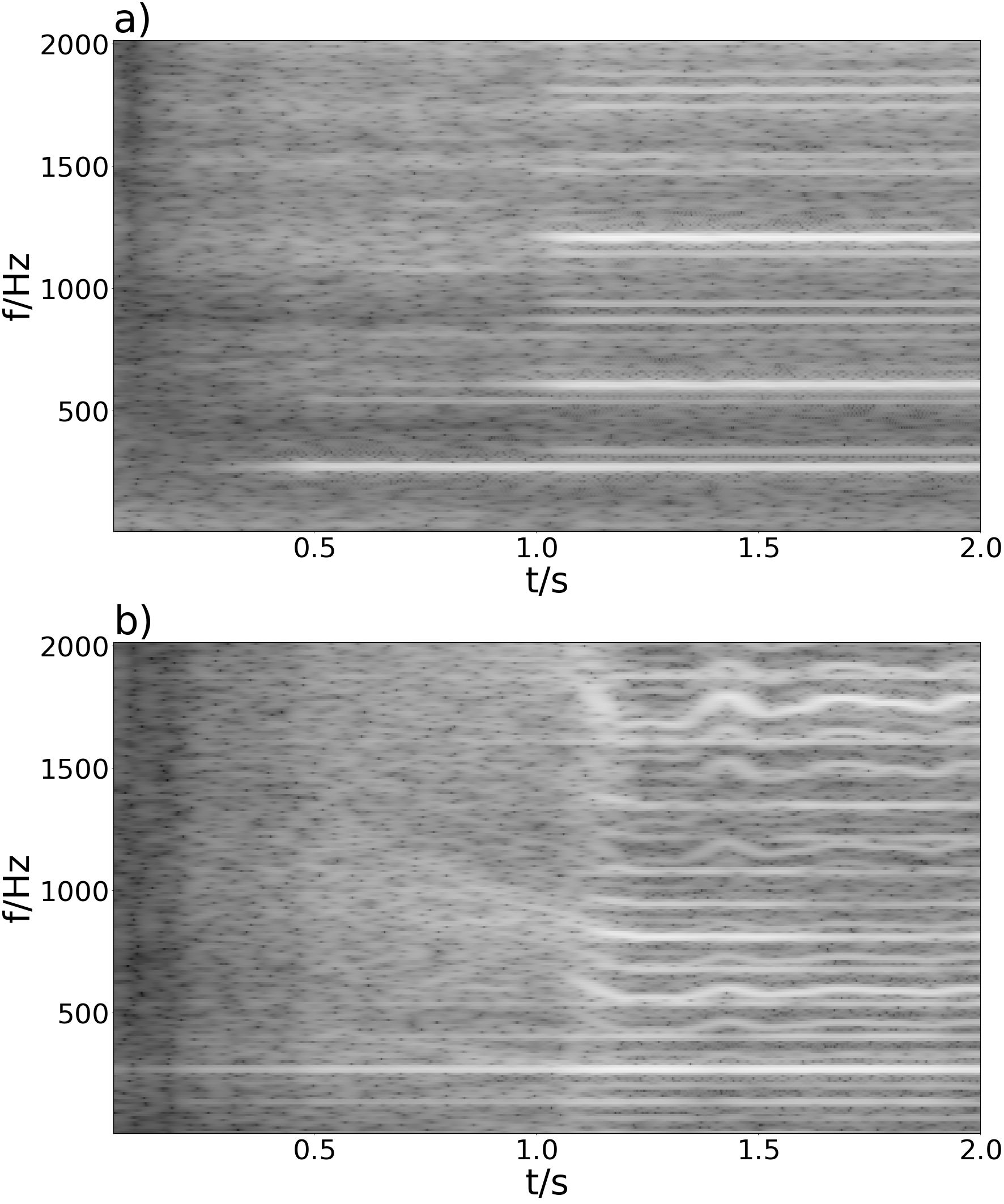}
    \caption{\label{fig:spectro} {Spectrogram of a) a recording of the multiphonic shown in Figure \ref{fig:MultiphonicFingering} and b) the synthesized version of the same multiphonic using the waveform of a recorded clarinet}}
  \end{center} 
\end{figure} 

Figure \ref{fig:spectro} shows spectrograms of a recorded sound (audio example 1 from \cite{Linke.2022} ), as well as a synthesized version, using the waveform of a clarinet sound (audio example 2 from \cite{Linke.2022} ). Both present a sudden transition from a rather noisy state with a weak fundamental frequency to bifurcations associated with multiphonics. Even though $\alpha$ was gradually changed, the fundamental frequency does not gradually split up. The additional frequencies appear suddenly, which also can be perceived when listening to the sounds.




At first glance, the synthesized sounds (audio example 2-4 from \cite{Linke.2022} ) sound different to the recording, when listening to them. However, these derivations occur due to spectral discrepancies of the assumed waveform. Nevertheless, the overall character is similar. A sudden transition from a noisy sound with one single, dominating harmonic series to the investigated interval of the multiphonic can be perceived similarly at the same point in time.

Comparing the three synthesized sounds to each other, there are strong perceivable spectral differences. Still, the overall character is very similar. So the transient behavior of the impulse pattern seems to have a more substantial impact than the actual spectral content.  





For a more realistic reproduction of the recording, the spectral content and the time series of $\alpha$ have to be chosen more carefully. Furthermore, adding more passed systems states $g_{k-}$ to the synthesis improves the noisy state.

\section{Conclusion}

The IPF can be used to model multiphonics on a clarinet, including associated transients from/into regular playing. Focussing on the overall system behavior in terms of stability, the IPF can model multiphonics that can be played at low blowing pressure on a clarinet, as shown for the Boehm system. Figure \ref{fig:BifBeta} indicates that this approach can be extended to high blowing pressures. The results are obtained at low computational cost, as just a logarithm has to be calculated for every investigated period $T$. The implementation is straightforward as every parameter of the model corresponds to one single property of the instrument (open finger hole). It seems reasonable that the method can be transferred to multiphonics with more than two dominating frequencies.

While searching the population of potentially playable multiphonics in the parametric space of the clarinet-tailored IPF, it could be shown that $\beta_2>\beta_1$ is crucial for producing multiphonics. A reflection point far away from the excitation point must be strong enough to disrupt periodic motion.

When modeling parameters are determined, selection criteria are helpful. The assumption that quasi-stable tone production involves a small derivative of the ratio $g_k / g_{k-}$ and the knowledge about the low blowing pressure during multiphonic production jointly constitute a useful criterion. When modeling multiphonics using the IPF, focussing on the correct ratio $g_k / g_{k-}$ alone is not sufficient.

As an example, in Figure \ref{fig:BifSimple} at $1/\alpha=2.2$, it might be possible to model a particular multiphonic. However, approaching this region by gradually changing $\alpha$, $1/\alpha \in [2, 2.4]$ results in multiphonics with a wrong frequency ratio, which leads to a perceivable pitch glide. Still calculating a likelihood for particular multiphonics, as shown in Section \ref{sec:multiCla}, allows for identifying regions of robust multiphonic production. There sudden transitions from chaotic states to desired multiphonics are present.

The results of the present investigation can directly be used to synthesize the related sounds. While the IPF yields the impulse pattern with the inherent pitch and dynamics, the spectral modeling follows subsequently and independently, thus leaving choices between realistic imitation and fancy instrumental mutation.

The synthesis method also seems to be capable of real-time transient modeling, as the calculation demand is very low.


\bibliography{refs}

\begin{thebibliography}{}

\bibitem [\protect \citeauthoryear {%
Argyris%
, Faust%
, Haase%
\BCBL {}\ \BBA {} Friedrich%
}{%
Argyris%
\ \protect \BOthers {.}}{%
{\protect \APACyear {2015}}%
}]{%
Argyris.2015}
\APACinsertmetastar {%
Argyris.2015}%
\begin{APACrefauthors}%
Argyris, J\BPBI H.%
, Faust, G.%
, Haase, M.%
\BCBL {}\ \BBA {} Friedrich, R.%
\end{APACrefauthors}%
\unskip\
\newblock
\APACrefYear{2015}.
\newblock
\APACrefbtitle {An Exploration of Dynamical Systems and Chaos: Completely
  Revised and Enlarged Second Edition} {An exploration of dynamical systems and
  chaos: Completely revised and enlarged second edition}.
\newblock
\APACaddressPublisher{Berlin, Heidelberg}{{Springer Berlin Heidelberg}}.
\newblock
\begin{APACrefDOI} \doi{10.1007/978-3-662-46042-9} \end{APACrefDOI}
\PrintBackRefs{\CurrentBib}

\bibitem [\protect \citeauthoryear {%
Backus%
}{%
Backus%
}{%
{\protect \APACyear {1963}}%
}]{%
Backus.1963}
\APACinsertmetastar {%
Backus.1963}%
\begin{APACrefauthors}%
Backus, J.%
\end{APACrefauthors}%
\unskip\
\newblock
\APACrefYearMonthDay{1963}{}{}.
\newblock
{\BBOQ}\APACrefatitle {Small--Vibration Theory of the Clarinet}
  {Small--vibration theory of the clarinet}.{\BBCQ}
\newblock
\APACjournalVolNumPages{The Journal of the Acoustical Society of
  America}{35}{3}{305--313}.
\newblock
\begin{APACrefDOI} \doi{10.1121/1.1918458} \end{APACrefDOI}
\PrintBackRefs{\CurrentBib}

\bibitem [\protect \citeauthoryear {%
Backus%
}{%
Backus%
}{%
{\protect \APACyear {1978}}%
}]{%
Backus.1978}
\APACinsertmetastar {%
Backus.1978}%
\begin{APACrefauthors}%
Backus, J.%
\end{APACrefauthors}%
\unskip\
\newblock
\APACrefYearMonthDay{1978}{}{}.
\newblock
{\BBOQ}\APACrefatitle {Multiphonic tones in the woodwind instruments}
  {Multiphonic tones in the woodwind instruments}.{\BBCQ}
\newblock
\APACjournalVolNumPages{The Journal of the Acoustical Society of
  America}{63}{2}{591--599}.
\newblock
\begin{APACrefDOI} \doi{10.1121/1.381758} \end{APACrefDOI}
\PrintBackRefs{\CurrentBib}

\bibitem [\protect \citeauthoryear {%
Bader%
}{%
Bader%
}{%
{\protect \APACyear {2008}}%
}]{%
Bader.2008}
\APACinsertmetastar {%
Bader.2008}%
\begin{APACrefauthors}%
Bader, R.%
\end{APACrefauthors}%
\unskip\
\newblock
\APACrefYearMonthDay{2008}{}{}.
\newblock
{\BBOQ}\APACrefatitle {Individual reed characteristics due to changed damping
  using coupled flow-structure and time-dependent geometry changing Finite-
  Element calculation} {Individual reed characteristics due to changed damping
  using coupled flow-structure and time-dependent geometry changing finite-
  element calculation}.{\BBCQ}
\newblock
\BIn{} \APACrefbtitle {Proceedings Forum Acusticum Joined with American
  Acoustical Society Paris} {Proceedings forum acusticum joined with american
  acoustical society paris}\ (\BPGS\ 3405--3410).
\newblock
\begin{APACrefURL}
  \url{http://webistem.com/acoustics2008/acoustics2008/cd1/data/articles/000441.pdf}
  \end{APACrefURL}
\PrintBackRefs{\CurrentBib}

\bibitem [\protect \citeauthoryear {%
Bader%
}{%
Bader%
}{%
{\protect \APACyear {2013}}%
}]{%
Bader.2013}
\APACinsertmetastar {%
Bader.2013}%
\begin{APACrefauthors}%
Bader, R.%
\end{APACrefauthors}%
\unskip\
\newblock
\APACrefYear{2013}.
\newblock
\APACrefbtitle {Nonlinearities and synchronization in musical acoustics and
  music psychology} {Nonlinearities and synchronization in musical acoustics
  and music psychology}\ (\BVOL~2).
\newblock
\APACaddressPublisher{Berlin [et al.]}{Springer}.
\newblock
\begin{APACrefDOI} \doi{10.1007/978-3-642-36098-5} \end{APACrefDOI}
\PrintBackRefs{\CurrentBib}

\bibitem [\protect \citeauthoryear {%
Bezanson%
, Edelman%
, Karpinski%
\BCBL {}\ \BBA {} Shah%
}{%
Bezanson%
\ \protect \BOthers {.}}{%
{\protect \APACyear {2017}}%
}]{%
Bezanson.2017}
\APACinsertmetastar {%
Bezanson.2017}%
\begin{APACrefauthors}%
Bezanson, J.%
, Edelman, A.%
, Karpinski, S.%
\BCBL {}\ \BBA {} Shah, V\BPBI B.%
\end{APACrefauthors}%
\unskip\
\newblock
\APACrefYearMonthDay{2017}{}{}.
\newblock
{\BBOQ}\APACrefatitle {Julia: A Fresh Approach to Numerical Computing} {Julia:
  A fresh approach to numerical computing}.{\BBCQ}
\newblock
\APACjournalVolNumPages{SIAM Review}{59}{1}{65--98}.
\newblock
\begin{APACrefDOI} \doi{10.1137/141000671} \end{APACrefDOI}
\PrintBackRefs{\CurrentBib}

\bibitem [\protect \citeauthoryear {%
Chen%
, Smith%
\BCBL {}\ \BBA {} Wolfe%
}{%
Chen%
\ \protect \BOthers {.}}{%
{\protect \APACyear {2011}}%
}]{%
Chen.2011}
\APACinsertmetastar {%
Chen.2011}%
\begin{APACrefauthors}%
Chen, J\BHBI M.%
, Smith, J.%
\BCBL {}\ \BBA {} Wolfe, J.%
\end{APACrefauthors}%
\unskip\
\newblock
\APACrefYearMonthDay{2011}{}{}.
\newblock
{\BBOQ}\APACrefatitle {Saxophonists tune vocal tract resonances in advanced
  performance techniques} {Saxophonists tune vocal tract resonances in advanced
  performance techniques}.{\BBCQ}
\newblock
\APACjournalVolNumPages{The Journal of the Acoustical Society of
  America}{129}{1}{415--426}.
\newblock
\begin{APACrefDOI} \doi{10.1121/1.3514423} \end{APACrefDOI}
\PrintBackRefs{\CurrentBib}

\bibitem [\protect \citeauthoryear {%
Cremer%
}{%
Cremer%
}{%
{\protect \APACyear {1984}}%
}]{%
Cremer.1984}
\APACinsertmetastar {%
Cremer.1984}%
\begin{APACrefauthors}%
Cremer, L.%
\end{APACrefauthors}%
\unskip\
\newblock
\APACrefYear{1984}.
\newblock
\APACrefbtitle {The physics of the violin} {The physics of the violin}.
\newblock
\APACaddressPublisher{Cambridge, Mass.}{{MIT Press}}.
\PrintBackRefs{\CurrentBib}

\bibitem [\protect \citeauthoryear {%
Dalmont%
, Gazengel%
, Gilbert%
\BCBL {}\ \BBA {} Kergomard%
}{%
Dalmont%
\ \protect \BOthers {.}}{%
{\protect \APACyear {1995}}%
}]{%
Dalmont.1995}
\APACinsertmetastar {%
Dalmont.1995}%
\begin{APACrefauthors}%
Dalmont, J\BPBI P.%
, Gazengel, B.%
, Gilbert, J.%
\BCBL {}\ \BBA {} Kergomard, J.%
\end{APACrefauthors}%
\unskip\
\newblock
\APACrefYearMonthDay{1995}{}{}.
\newblock
{\BBOQ}\APACrefatitle {Some aspects of tuning and clean intonation in reed
  instruments} {Some aspects of tuning and clean intonation in reed
  instruments}.{\BBCQ}
\newblock
\APACjournalVolNumPages{Applied Acoustics}{46}{1}{19--60}.
\newblock
\begin{APACrefDOI} \doi{10.1016/0003-682X(95)93950-M} \end{APACrefDOI}
\PrintBackRefs{\CurrentBib}

\bibitem [\protect \citeauthoryear {%
Dalmont%
, Gilbert%
\BCBL {}\ \BBA {} Ollivier%
}{%
Dalmont%
\ \protect \BOthers {.}}{%
{\protect \APACyear {2003}}%
}]{%
Dalmont.2003}
\APACinsertmetastar {%
Dalmont.2003}%
\begin{APACrefauthors}%
Dalmont, J\BHBI P.%
, Gilbert, J.%
\BCBL {}\ \BBA {} Ollivier, S.%
\end{APACrefauthors}%
\unskip\
\newblock
\APACrefYearMonthDay{2003}{}{}.
\newblock
{\BBOQ}\APACrefatitle {Nonlinear characteristics of single-reed instruments:
  quasistatic volume flow and reed opening measurements} {Nonlinear
  characteristics of single-reed instruments: quasistatic volume flow and reed
  opening measurements}.{\BBCQ}
\newblock
\APACjournalVolNumPages{The Journal of the Acoustical Society of
  America}{114}{4 Pt 1}{2253--2262}.
\newblock
\begin{APACrefDOI} \doi{10.1121/1.1603235} \end{APACrefDOI}
\PrintBackRefs{\CurrentBib}

\bibitem [\protect \citeauthoryear {%
Datseris%
}{%
Datseris%
}{%
{\protect \APACyear {2018}}%
}]{%
Datseris.2018}
\APACinsertmetastar {%
Datseris.2018}%
\begin{APACrefauthors}%
Datseris, G.%
\end{APACrefauthors}%
\unskip\
\newblock
\APACrefYearMonthDay{2018}{}{}.
\newblock
{\BBOQ}\APACrefatitle {DynamicalSystems.jl: A Julia software library for chaos
  and nonlinear dynamics} {Dynamicalsystems.jl: A julia software library for
  chaos and nonlinear dynamics}.{\BBCQ}
\newblock
\APACjournalVolNumPages{Journal of Open Source Software}{3}{23}{598}.
\newblock
\begin{APACrefDOI} \doi{10.21105/joss.00598} \end{APACrefDOI}
\PrintBackRefs{\CurrentBib}

\bibitem [\protect \citeauthoryear {%
Fabre%
, Gilbert%
\BCBL {}\ \BBA {} Hirschberg%
}{%
Fabre%
\ \protect \BOthers {.}}{%
{\protect \APACyear {2018}}%
}]{%
Fabre.2018}
\APACinsertmetastar {%
Fabre.2018}%
\begin{APACrefauthors}%
Fabre, B.%
, Gilbert, J.%
\BCBL {}\ \BBA {} Hirschberg, A.%
\end{APACrefauthors}%
\unskip\
\newblock
\APACrefYearMonthDay{2018}{}{}.
\newblock
{\BBOQ}\APACrefatitle {Modeling of Wind Instruments} {Modeling of wind
  instruments}.{\BBCQ}
\newblock
\BIn{} R.~Bader\ (\BED), \APACrefbtitle {Springer Handbook of Systematic
  Musicology} {Springer handbook of systematic musicology}\ (\BPGS\ 121--139).
\newblock
\APACaddressPublisher{Berlin, Heidelberg}{Springer}.
\newblock
\begin{APACrefDOI} \doi{10.1007/978-3-662-55004-5_7} \end{APACrefDOI}
\PrintBackRefs{\CurrentBib}

\bibitem [\protect \citeauthoryear {%
Farmer%
}{%
Farmer%
}{%
{\protect \APACyear {1977}}%
}]{%
Farmer.1977}
\APACinsertmetastar {%
Farmer.1977}%
\begin{APACrefauthors}%
Farmer, G\BPBI J.%
\end{APACrefauthors}%
\unskip\
\newblock
\APACrefYear{1977}.
\unskip\
\newblock
\APACrefbtitle {Multiphonic trills and tremolos for clarinet} {Multiphonic
  trills and tremolos for clarinet}\ \APACtypeAddressSchool {\BUPhD}{}{}.
\unskip\
\newblock
\APACaddressSchool {Ann Arbor, Mich.}{{@Eugene, Or., Univ. of Oregon, Diss.,}}.
\PrintBackRefs{\CurrentBib}

\bibitem [\protect \citeauthoryear {%
Fletcher%
}{%
Fletcher%
}{%
{\protect \APACyear {1978}}%
}]{%
Fletcher.1978}
\APACinsertmetastar {%
Fletcher.1978}%
\begin{APACrefauthors}%
Fletcher, N\BPBI H.%
\end{APACrefauthors}%
\unskip\
\newblock
\APACrefYearMonthDay{1978}{}{}.
\newblock
{\BBOQ}\APACrefatitle {Mode locking in nonlinearly excited inharmonic musical
  oscillators} {Mode locking in nonlinearly excited inharmonic musical
  oscillators}.{\BBCQ}
\newblock
\APACjournalVolNumPages{The Journal of the Acoustical Society of
  America}{64}{6}{1566--1569}.
\newblock
\begin{APACrefDOI} \doi{10.1121/1.382139} \end{APACrefDOI}
\PrintBackRefs{\CurrentBib}

\bibitem [\protect \citeauthoryear {%
Fletcher%
}{%
Fletcher%
}{%
{\protect \APACyear {1979}}%
}]{%
Fletcher.1979}
\APACinsertmetastar {%
Fletcher.1979}%
\begin{APACrefauthors}%
Fletcher, N\BPBI H.%
\end{APACrefauthors}%
\unskip\
\newblock
\APACrefYearMonthDay{1979}{}{}.
\newblock
{\BBOQ}\APACrefatitle {Air flow and sound generation in musical wind
  instruments} {Air flow and sound generation in musical wind
  instruments}.{\BBCQ}
\newblock
\APACjournalVolNumPages{Annual Review of Fluid Mechanics}{11}{1}{123--146}.
\newblock
\begin{APACrefDOI} \doi{10.1146/annurev.fl.11.010179.001011),} \end{APACrefDOI}
\PrintBackRefs{\CurrentBib}

\bibitem [\protect \citeauthoryear {%
Fletcher%
\ \BBA {} Rossing%
}{%
Fletcher%
\ \BBA {} Rossing%
}{%
{\protect \APACyear {2010}}%
}]{%
Fletcher.2010}
\APACinsertmetastar {%
Fletcher.2010}%
\begin{APACrefauthors}%
Fletcher, N\BPBI H.%
\BCBT {}\ \BBA {} Rossing, T\BPBI D.%
\end{APACrefauthors}%
\unskip\
\newblock
\APACrefYear{2010}.
\newblock
\APACrefbtitle {The physics of musical instruments} {The physics of musical
  instruments}\ (\PrintOrdinal{2. ed., [rpt..]}\ \BEd).
\newblock
\APACaddressPublisher{New York, NY}{Springer}.
\newblock
\begin{APACrefDOI} \doi{10.1007/978-0-387-21603-4} \end{APACrefDOI}
\PrintBackRefs{\CurrentBib}

\bibitem [\protect \citeauthoryear {%
Giordano%
}{%
Giordano%
}{%
{\protect \APACyear {2018}}%
}]{%
Giordano.2018}
\APACinsertmetastar {%
Giordano.2018}%
\begin{APACrefauthors}%
Giordano, N.%
\end{APACrefauthors}%
\unskip\
\newblock
\APACrefYearMonthDay{2018}{}{}.
\newblock
{\BBOQ}\APACrefatitle {Some Observations on the Physics of Stringed
  Instruments} {Some observations on the physics of stringed
  instruments}.{\BBCQ}
\newblock
\BIn{} R.~Bader\ (\BED), \APACrefbtitle {Springer Handbook of Systematic
  Musicology} {Springer handbook of systematic musicology}\ (\BPGS\ 105--119).
\newblock
\APACaddressPublisher{Berlin, Heidelberg}{Springer}.
\newblock
\begin{APACrefDOI} \doi{10.1007/978-3-662-55004-5_6} \end{APACrefDOI}
\PrintBackRefs{\CurrentBib}

\bibitem [\protect \citeauthoryear {%
Hirschberg%
, Pelorson%
\BCBL {}\ \BBA {} Gilbert%
}{%
Hirschberg%
\ \protect \BOthers {.}}{%
{\protect \APACyear {1996}}%
}]{%
Hirschberg.1996}
\APACinsertmetastar {%
Hirschberg.1996}%
\begin{APACrefauthors}%
Hirschberg, A.%
, Pelorson, X.%
\BCBL {}\ \BBA {} Gilbert, J.%
\end{APACrefauthors}%
\unskip\
\newblock
\APACrefYearMonthDay{1996}{}{}.
\newblock
{\BBOQ}\APACrefatitle {Aeroacoustics of musical instruments} {Aeroacoustics of
  musical instruments}.{\BBCQ}
\newblock
\APACjournalVolNumPages{Meccanica}{31}{2}{131--141}.
\newblock
\begin{APACrefDOI} \doi{10.1007/BF00426256} \end{APACrefDOI}
\PrintBackRefs{\CurrentBib}

\bibitem [\protect \citeauthoryear {%
Keefe%
}{%
Keefe%
}{%
{\protect \APACyear {1990}}%
}]{%
Keefe.1990}
\APACinsertmetastar {%
Keefe.1990}%
\begin{APACrefauthors}%
Keefe, D\BPBI H.%
\end{APACrefauthors}%
\unskip\
\newblock
\APACrefYearMonthDay{1990}{}{}.
\newblock
{\BBOQ}\APACrefatitle {Woodwind air column models} {Woodwind air column
  models}.{\BBCQ}
\newblock
\APACjournalVolNumPages{The Journal of the Acoustical Society of
  America}{88}{1}{35--51}.
\newblock
\begin{APACrefDOI} \doi{10.1121/1.399911} \end{APACrefDOI}
\PrintBackRefs{\CurrentBib}

\bibitem [\protect \citeauthoryear {%
Keefe%
\ \BBA {} Laden%
}{%
Keefe%
\ \BBA {} Laden%
}{%
{\protect \APACyear {1991}}%
}]{%
Keefe.1991}
\APACinsertmetastar {%
Keefe.1991}%
\begin{APACrefauthors}%
Keefe, D\BPBI H.%
\BCBT {}\ \BBA {} Laden, B.%
\end{APACrefauthors}%
\unskip\
\newblock
\APACrefYearMonthDay{1991}{}{}.
\newblock
{\BBOQ}\APACrefatitle {Correlation dimension of woodwind multiphonic tones}
  {Correlation dimension of woodwind multiphonic tones}.{\BBCQ}
\newblock
\APACjournalVolNumPages{The Journal of the Acoustical Society of
  America}{90}{4}{1754--1765}.
\newblock
\begin{APACrefDOI} \doi{10.1121/1.401656} \end{APACrefDOI}
\PrintBackRefs{\CurrentBib}

\bibitem [\protect \citeauthoryear {%
Linke%
, Bader%
\BCBL {}\ \BBA {} Mores%
}{%
Linke%
\ \protect \BOthers {.}}{%
{\protect \APACyear {2019}}%
{\protect \APACexlab {{\protect \BCnt {1}}}}}]{%
Linke.2019b}
\APACinsertmetastar {%
Linke.2019b}%
\begin{APACrefauthors}%
Linke, S.%
, Bader, R.%
\BCBL {}\ \BBA {} Mores, R.%
\end{APACrefauthors}%
\unskip\
\newblock
\APACrefYearMonthDay{2019{\protect \BCnt {1}}}{}{}.
\newblock
{\BBOQ}\APACrefatitle {The impulse pattern formulation (IPF) as a model of
  musical instruments---Investigation of stability and limits} {The impulse
  pattern formulation (ipf) as a model of musical instruments---investigation
  of stability and limits}.{\BBCQ}
\newblock
\APACjournalVolNumPages{Chaos: An Interdisciplinary Journal of Nonlinear
  Science}{29}{10}{103109}.
\newblock
\begin{APACrefDOI} \doi{10.1063/1.5092511} \end{APACrefDOI}
\PrintBackRefs{\CurrentBib}

\bibitem [\protect \citeauthoryear {%
Linke%
, Bader%
\BCBL {}\ \BBA {} Mores%
}{%
Linke%
\ \protect \BOthers {.}}{%
{\protect \APACyear {2019}}%
{\protect \APACexlab {{\protect \BCnt {2}}}}}]{%
Linke.2019c}
\APACinsertmetastar {%
Linke.2019c}%
\begin{APACrefauthors}%
Linke, S.%
, Bader, R.%
\BCBL {}\ \BBA {} Mores, R.%
\end{APACrefauthors}%
\unskip\
\newblock
\APACrefYearMonthDay{2019{\protect \BCnt {2}}}{}{}.
\newblock
{\BBOQ}\APACrefatitle {The Impulse Pattern Formulation (IPF) as a nonlinear
  model of musical instruments} {The impulse pattern formulation (ipf) as a
  nonlinear model of musical instruments}.{\BBCQ}
\newblock
\BIn{} M.~Kob\ (\BED), \APACrefbtitle {Proceedings of the International
  Symposium on Music Acoustics 2019 - ISMA 2019} {Proceedings of the
  international symposium on music acoustics 2019 - isma 2019}\ (\BPGS\
  336--345).
\newblock
\APACaddressPublisher{Berlin}{}.
\newblock
\begin{APACrefURL}
  \url{http://pub.dega-akustik.de/ISMA2019/data/ISMA_proceedings_all.pdf}
  \end{APACrefURL}
\PrintBackRefs{\CurrentBib}

\bibitem [\protect \citeauthoryear {%
Linke%
, Bader%
\BCBL {}\ \BBA {} Mores%
}{%
Linke%
\ \protect \BOthers {.}}{%
{\protect \APACyear {2021}}%
}]{%
Linke.2021c}
\APACinsertmetastar {%
Linke.2021c}%
\begin{APACrefauthors}%
Linke, S.%
, Bader, R.%
\BCBL {}\ \BBA {} Mores, R.%
\end{APACrefauthors}%
\unskip\
\newblock
\APACrefYearMonthDay{2021}{}{}.
\newblock
\APACrefbtitle {Modeling synchronization in human musical rhythms using Impulse
  Pattern Formulation (IPF).} {Modeling synchronization in human musical
  rhythms using impulse pattern formulation (ipf).}
\newblock
\begin{APACrefURL} \url{http://arxiv.org/pdf/2112.03218v1} \end{APACrefURL}
\PrintBackRefs{\CurrentBib}

\bibitem [\protect \citeauthoryear {%
Linke%
, Bader%
\BCBL {}\ \BBA {} Mores%
}{%
Linke%
\ \protect \BOthers {.}}{%
{\protect \APACyear {2022}}%
}]{%
Linke.2022}
\APACinsertmetastar {%
Linke.2022}%
\begin{APACrefauthors}%
Linke, S.%
, Bader, R.%
\BCBL {}\ \BBA {} Mores, R.%
\end{APACrefauthors}%
\unskip\
\newblock
\APACrefYearMonthDay{2022}{}{}.
\newblock
\APACrefbtitle {Multiphonic of clarinet synthesized using Impulse Pattern
  Formulation (IPF).} {Multiphonic of clarinet synthesized using impulse
  pattern formulation (ipf).}
\newblock
\APACaddressPublisher{}{Zenodo}.
\newblock
\begin{APACrefURL} \url{https://zenodo.org/record/5849430} \end{APACrefURL}
\newblock
\begin{APACrefDOI} \doi{10.5281/ZENODO.5849430} \end{APACrefDOI}
\PrintBackRefs{\CurrentBib}

\bibitem [\protect \citeauthoryear {%
Maganza%
, Causs{\'e}%
\BCBL {}\ \BBA {} Lalo{\"e}%
}{%
Maganza%
\ \protect \BOthers {.}}{%
{\protect \APACyear {1986}}%
}]{%
Maganza.1986}
\APACinsertmetastar {%
Maganza.1986}%
\begin{APACrefauthors}%
Maganza, C.%
, Causs{\'e}, R.%
\BCBL {}\ \BBA {} Lalo{\"e}, F.%
\end{APACrefauthors}%
\unskip\
\newblock
\APACrefYearMonthDay{1986}{}{}.
\newblock
{\BBOQ}\APACrefatitle {Bifurcations, period doublings and chaos in clarinetlike
  systems} {Bifurcations, period doublings and chaos in clarinetlike
  systems}.{\BBCQ}
\newblock
\APACjournalVolNumPages{EPL (Europhysics Letters)}{1}{6}{295--302}.
\newblock
\begin{APACrefDOI} \doi{10.1209/0295-5075/1/6/005} \end{APACrefDOI}
\PrintBackRefs{\CurrentBib}

\bibitem [\protect \citeauthoryear {%
McIntyre%
, Schumacher%
\BCBL {}\ \BBA {} Woodhouse%
}{%
McIntyre%
\ \protect \BOthers {.}}{%
{\protect \APACyear {1983}}%
}]{%
McIntyre.1983}
\APACinsertmetastar {%
McIntyre.1983}%
\begin{APACrefauthors}%
McIntyre, M\BPBI E.%
, Schumacher, R\BPBI T.%
\BCBL {}\ \BBA {} Woodhouse, J.%
\end{APACrefauthors}%
\unskip\
\newblock
\APACrefYearMonthDay{1983}{}{}.
\newblock
{\BBOQ}\APACrefatitle {On the oscillations of musical instruments} {On the
  oscillations of musical instruments}.{\BBCQ}
\newblock
\APACjournalVolNumPages{The Journal of the Acoustical Society of
  America}{74}{5}{1325--1345}.
\newblock
\begin{APACrefDOI} \doi{10.1121/1.390157} \end{APACrefDOI}
\PrintBackRefs{\CurrentBib}

\bibitem [\protect \citeauthoryear {%
Nederveen%
}{%
Nederveen%
}{%
{\protect \APACyear {1969}}%
}]{%
Nederveen.1969}
\APACinsertmetastar {%
Nederveen.1969}%
\begin{APACrefauthors}%
Nederveen, C\BPBI J.%
\end{APACrefauthors}%
\unskip\
\newblock
\APACrefYear{1969}.
\unskip\
\newblock
\APACrefbtitle {Acoustical aspects of woodwind instruments} {Acoustical aspects
  of woodwind instruments}\ \APACtypeAddressSchool {\BPhD}{Delft}{{TECHNISCHE
  HOGESCHOOL DELFT}}.
\unskip\
\newblock
\begin{APACrefURL}
  \url{https://repository.tudelft.nl/islandora/object/uuid:01b56232-d1c8-4394-902d-e5e51b9ec223}
  \end{APACrefURL}
\PrintBackRefs{\CurrentBib}

\bibitem [\protect \citeauthoryear {%
Roche%
}{%
Roche%
}{%
{\protect \APACyear {2014}}%
}]{%
Roche.2014}
\APACinsertmetastar {%
Roche.2014}%
\begin{APACrefauthors}%
Roche, H.%
\end{APACrefauthors}%
\unskip\
\newblock
\APACrefYearMonthDay{2014}{}{}.
\newblock
\APACrefbtitle {$\ldots$on close dyad multiphonics for Bb clarinet.}
  {$\ldots$on close dyad multiphonics for bb clarinet.}
\newblock
\begin{APACrefURL}
  [{2022-01-14}]\url{https://heatherroche.net/2014/07/02/on-close-dyad-multiphonics-for-bb-clarinet/}
  \end{APACrefURL}
\PrintBackRefs{\CurrentBib}

\bibitem [\protect \citeauthoryear {%
Roche%
}{%
Roche%
}{%
{\protect \APACyear {2018}}%
}]{%
Roche.2018}
\APACinsertmetastar {%
Roche.2018}%
\begin{APACrefauthors}%
Roche, H.%
\end{APACrefauthors}%
\unskip\
\newblock
\APACrefYearMonthDay{2018}{}{}.
\newblock
\APACrefbtitle {27 Easy Bb Clarinet Multiphonics.} {27 easy bb clarinet
  multiphonics.}
\newblock
\begin{APACrefURL}
  [{2022-01-14}]\url{https://heatherroche.net/2018/09/13/27-easy-bb-clarinet-multiphonics/}
  \end{APACrefURL}
\PrintBackRefs{\CurrentBib}

\bibitem [\protect \citeauthoryear {%
Rodet%
\ \BBA {} Vergez%
}{%
Rodet%
\ \BBA {} Vergez%
}{%
{\protect \APACyear {1999}}%
{\protect \APACexlab {{\protect \BCnt {1}}}}}]{%
Rodet.1999b}
\APACinsertmetastar {%
Rodet.1999b}%
\begin{APACrefauthors}%
Rodet, X.%
\BCBT {}\ \BBA {} Vergez, C.%
\end{APACrefauthors}%
\unskip\
\newblock
\APACrefYearMonthDay{1999{\protect \BCnt {1}}}{}{}.
\newblock
{\BBOQ}\APACrefatitle {Nonlinear Dynamics in Physical Models: From Basic Models
  to True Musical-Instrument Models} {Nonlinear dynamics in physical models:
  From basic models to true musical-instrument models}.{\BBCQ}
\newblock
\APACjournalVolNumPages{Computer Music Journal}{23}{3}{35--49}.
\newblock
\begin{APACrefDOI} \doi{10.1162/014892699559878} \end{APACrefDOI}
\PrintBackRefs{\CurrentBib}

\bibitem [\protect \citeauthoryear {%
Rodet%
\ \BBA {} Vergez%
}{%
Rodet%
\ \BBA {} Vergez%
}{%
{\protect \APACyear {1999}}%
{\protect \APACexlab {{\protect \BCnt {2}}}}}]{%
Rodet.1999}
\APACinsertmetastar {%
Rodet.1999}%
\begin{APACrefauthors}%
Rodet, X.%
\BCBT {}\ \BBA {} Vergez, C.%
\end{APACrefauthors}%
\unskip\
\newblock
\APACrefYearMonthDay{1999{\protect \BCnt {2}}}{}{}.
\newblock
{\BBOQ}\APACrefatitle {Nonlinear Dynamics in Physical Models: Simple
  Feedback-Loop Systems and Properties} {Nonlinear dynamics in physical models:
  Simple feedback-loop systems and properties}.{\BBCQ}
\newblock
\APACjournalVolNumPages{Computer Music Journal}{23}{3}{18--34}.
\newblock
\begin{APACrefDOI} \doi{10.1162/014892699559869} \end{APACrefDOI}
\PrintBackRefs{\CurrentBib}

\bibitem [\protect \citeauthoryear {%
Taillard%
, Kergomard%
\BCBL {}\ \BBA {} Lalo{\"e}%
}{%
Taillard%
\ \protect \BOthers {.}}{%
{\protect \APACyear {2010}}%
}]{%
Taillard.2010}
\APACinsertmetastar {%
Taillard.2010}%
\begin{APACrefauthors}%
Taillard, P\BHBI A.%
, Kergomard, J.%
\BCBL {}\ \BBA {} Lalo{\"e}, F.%
\end{APACrefauthors}%
\unskip\
\newblock
\APACrefYearMonthDay{2010}{}{}.
\newblock
{\BBOQ}\APACrefatitle {Iterated maps for clarinet-like systems} {Iterated maps
  for clarinet-like systems}.{\BBCQ}
\newblock
\APACjournalVolNumPages{Nonlinear Dynamics}{62}{1}{253--271}.
\newblock
\begin{APACrefDOI} \doi{10.1007/s11071-010-9715-5} \end{APACrefDOI}
\PrintBackRefs{\CurrentBib}

\end{thebibliography}

\end{document}